\documentclass{emulateapj}

\def\se#1{\S\ref{sec:#1}}
\def \kms{\ifmmode\,{\rm km}\,{\rm s}^{-1}\else km$\,$s$^{-1}$\fi}

\newcommand{\myemail}{lauren@physics.ubc.ca}
\newcommand{\etal}{et al.\@}
\newcommand{\eg}{e.g.\@}
\newcommand{\ie}{i.e.\@}
\newcommand{\chisqr}{$\chi^2$}

\slugcomment{For Publication in ApJ Supplements}

\shorttitle{Structure of Disk Dominated Galaxies: II. Color Gradients}
\shortauthors{MacArthur, Courteau, Bell, \& Holtzman}

\begin{document}

\title{Structure of Disk Dominated Galaxies: II. Color Gradients and 
	Stellar Population Models}

\author{Lauren A. MacArthur}
\affil{Department of Physics \& Astronomy, University of British Columbia,
    6224 Agricultural Road, Vancouver, BC CANADA V6T 1Z1}
\email{\myemail}

\author{St\'{e}phane Courteau}
\affil{Department of Physics, Queen's University, Kingston, ON CANADA K7L 3N6}
\email{courteau@astro.queensu.ca}

\author{Eric Bell}
\affil{Max-Planck-Institut f\"{u}r Astronomie, K\"{o}nigstuhl 17,
     D-69117 Heidelberg, Germany}
\email{bell@mpia-hd.mpg.de}

\and
\author{Jon A.~Holtzman}
\affil{Department of Astronomy, New Mexico State University,
     Box 30001, Department 45000, Las Cruces, NM 88003-8001}
\email{holtz@astro.nmsu.edu}

%\altaffiltext{1}{Visiting Astronomer at Lowell and Kitt Peak National
%   Observatory. KPNO is operated by AURA, Inc. under
%   cooperative agreement with the National Science Foundation.}

\begin{abstract}

We investigate optical and near-IR color gradients in a sample of 172 
low-inclination galaxies spanning Hubble types S0--Irr.  The colors are 
compared to stellar population synthesis models from which luminosity-weighted
average ages and metallicities are determined.  We explore the effects of 
different underlying star formation histories and additional bursts of 
star formation.  Our results are robust in a relative sense
under the assumption that our galaxies shared a similar underlying star 
formation history and that no bursts involving more than $\sim$10\% of the 
galaxy mass have occurred in the past 1--2 Gyr.
Because the observed gradients show radial structure, we measure ``inner'' 
and ``outer'' disk age and metallicity gradients.
Trends in age and metallicity and their gradients are explored as a function 
of Hubble type, rotational velocity, total near-IR galaxy magnitude, central 
surface brightness, and scale length.  We find strong correlations in age and 
metallicity with Hubble-type, rotational velocity, total magnitude, and 
central surface brightness in the sense that earlier-type, faster rotating, 
more luminous, and higher surface brightness galaxies are older and more 
metal-rich, suggesting an early and more rapid
star formation history for these galaxies.  The increasing trends with 
rotational velocity and total magnitude level off for 
$V_{\mathrm{rot}} \ga 120$ \kms\ and M$_{K} \la -23$ mag respectively.  This 
effect is stronger for metallicity (than age) 
which could reflect a threshold potential above which all metals are retained
and thus metallicity saturates at the yield.  Outer disk gradients are found 
to be weaker than the inner gradients as expected for a slower variation of 
the potential and surface brightness in the outer parts.  We find that stronger
age gradients are associated with weaker metallicity gradients.
Trends in gradients with galaxy parameters are compared with model 
predictions: these trends do not agree with predictions of semi-analytic 
models of hierarchical galaxy formation, possibly due to the effect of 
bar-induced radial flows.  The observed 
trends are in agreement with chemo-spectrophotometric models of spiral galaxy 
evolution based on CDM-motivated scaling laws but none of the hierarchical 
merging characteristics, implying a strong dependence of the star formation 
history of spiral galaxies on the galaxy potential and halo spin parameter.

\end{abstract}

%% Keywords should appear after the \end{abstract} command. The uncommented
%% example has been keyed in ApJ style. See the instructions to authors
%% for the journal to which you are submitting your paper to determine
%% what keyword punctuation is appropriate.


\section{Introduction}

Color gradients in galaxies reveal information about the nature of their 
stellar populations via age and metallicity trends, and the 
amount and distribution of dust. 
The technique of using broad-band colors as a probe of the stellar 
populations and star formation histories (SFH) of galaxies, 
pioneered by, \eg\, Searle, Sargent, \& Bagnuolo (1973) and 
Tinsley \& Gunn (1976), is still reflected in the modern studies 
of Peletier \& Balcells (1994), de~Jong (1996), and Bell \& de~Jong (2000; 
hereafter BdJ00).  Early analyses, however, suffered significantly from 
degeneracies between age and metallicity.
%significant degeneracies
%to separate age from metallicity effects.  
Worthey (1994) quantified the age-metallicity degeneracy that 
exists in optical broad-band colors as $\Delta\mathrm{age}/\Delta{Z} \sim 3/2$.
This implies that the composite spectrum of an old stellar population is nearly
indistinguishable from that of a younger but more metal-rich population (and 
vice versa).  This degeneracy can be partially broken with infrared 
photometry (\eg\ $H$ or $K$ band) in addition to optical colors 
(de~Jong 1996).  Cardiel \etal\ (2003) have quantified the relative ability 
of different color and absorption-line index combinations to constrain
physical parameters of composite stellar populations.
% by defining a ``suitability parameter''.  According to this parameter, 
Their results find that inclusion of an 
infrared band improves the predictive power of the stellar population 
diagnostics by $\sim 30$x over using optical colors alone.  The 
interpretation of broad-band color gradients also relies on a careful mapping 
of the dust extinction within a galaxy (Witt, Thronson, \& Capuano 1992; 
de~Jong 1996).  
While dust opacity is much reduced at redder wavelengths, its effects may 
still be non-negligible in the determination of the stellar content of 
late-type spiral galaxies and must be considered in the final interpretation
of color gradients.  

A successful galaxy formation theory must determine the formation timescales 
for bulges and disks and reproduce the observed stellar ages and chemical 
distributions in galaxies as a function of galaxy properties.  
Competing bulge/disk formation scenarios entail distinct 
structural and kinematic signatures, star formation rates (SFRs), and mixing 
of the bulge and disk material, that a study of age and metallicity 
indicators could help disentangle.  Current theories for the formation of disk 
bulges generally fall under two basic pictures: monolithic collapse (\eg\ 
Larson 1974; Calberg 1984) and hierarchical merging (\eg\ Kauffmann 1996; 
Cole \etal\ 2000).
The monolithic dissipative collapse of a gas cloud at high redshift is 
consistent with strong {\it bulge} metallicity 
gradients (on the order of 0.5 dex/decade within the central $\sim$2 kpc, but
flattening in the outer parts (Larson 1974)), with metallicity 
decreasing with increasing radius, shallow positive age gradients 
(the centers are slightly younger than the outer parts), 
and old stellar populations.  In this ``monolithic'' scenario the disk would
accrete onto the already formed bulge, and thus be younger than the bulge.
The dominant signatures in the hierarchical scenario depend on the time 
evolution of the merger
rate: if mergers were most frequent at high redshift, as predicted by 
present-epoch cosmological models, signatures should be similar to those of 
dissipative collapse.  More recent mergers and tidal disruptions would result 
in flatter metallicity gradients, and a significant fraction of young or 
intermediate-age stars.  In either of the above two scenarios, 
secular evolution can also drive disk gas and stars to the center via bar 
instabilities.  In general, this scenario predicts that bulges will be similar 
in metallicity and age and structurally linked to disks 
(Courteau, de Jong, \& Broeils 1996; Courteau 1996).  
Note, however, that the number of bar formation - gas inflow - star 
formation - bar dissolution cycles that can occur in a galaxy may be limited, 
as the central mass formed in the first cycle can prevent subsequent formation 
of a bar Wyse (1999). \footnote{Only progressively larger bars in the centers 
of exponential bulges would be allowed to form in a recurring scenario as a 
result of the disrupting dynamical effect of a growing nucleus 
(H.-W. Rix 1999, private communication, as reported in Carollo 1999).}
Radial migration of disk material (stars and gas) can also be induced by
non-axisymmetric spiral waves (Sellwood \& Binney 2002), which would 
necessarilly flatten metallicity gradients in disks.

Current observational results (O/H gradients, SFRs, constant IMF, gas 
distribution) favor an inside-out formation for the 
Milky Way, but the details about age and metallicity gradients in our 
own Galaxy are conflicting (Freeman \& Bland-Hawthorn 2002).
Integrated studies of stellar populations in external galaxies can 
restrict the range of possible interpretations, but formation 
timescales and chemical evolution of external spiral galaxies have barely 
been explored (\eg\ Matteucci 2002).  These issues have become even more 
important now that 
cosmologically-motivated numerical simulations are capable of 
resolving the formation of disks and bulges.  Furthermore, semi-analytic 
galaxy evolution models use recipes to describe bulge and disk evolution, 
but if different types of galaxies have different bulge formation mechanisms, 
these models may give misleading results (\eg\ Somerville \& Primack 
1999; van den Bosch 2000), unless a variety of bulge formation mechanisms 
are modeled (\eg\ Cole \etal\ 2000). 

Existing studies of spiral bulges and disks have placed tentative constraints 
on the source of their color gradients and which galaxy parameters most 
strongly correlate with SFHs and, hence, on specific formation scenarios, but 
some of the results are conflicting.  In an study of color gradients in 
early-type bulges ($\le$Sb) using HST and ground-based data, 
Peletier \etal\ (1999) infered that early-type bulges are old (absolute ages 
not well determined) with a small age spread ($\la$ 2 Gyr) for all early-type 
bulges measured at the bulge effective radius.  The old and narrow age range 
makes it unlikely that these bulges would have formed via secular evolution.
The same conclusion does not hold for their (few) later-type galaxies.
Peletier \etal\ (1999) conclude that the intrinsic color gradients of  
early-type bulges are caused mainly by metallicity gradients (in agreement 
with other similar studies; \eg\ Mehlert et al. 2003), consistent with the 
monolithic collapse scenario.  However, based on the lack of $r^{1/4}$ shaped 
bulges found in a later study of bulge SB profiles of S0--Sbc galaxies
using HST near-IR high resolution imaging, Balcells \etal\ (2003) have 
revised their interpretation and conclude that these bulges could not have 
formed from violent relaxation in mergers, but rather are more consistent 
with secular formation processes.  On the other hand, in an HST study of 
spatially resolved colors of high 
redshift ($\bar{z} \sim 0.5$) galaxies in the Hubble Deep Field, 
Abraham \etal\ (1999) rule out metallicity 
gradients as major contributors to galaxy color distributions.  They do find,
in agreement with Peletier \etal\ (1999), that large bulges are 
significantly older than their disks and therefore rule out secular evolution 
formation processes in favor of a gradual disk formation by accretion of gas. 
However, the flat metallicity distribution of the bulge is not easily 
explained in this scenario (particularly since the lack of ``$r^{1/4}$'' 
bulges rules out major merging as a cause of gradient flattening).

The most comprehensive study to date of color gradients in disk galaxies is 
that of BdJ00, who compute age and metallicity gradients for a sample of 121
nearby low-inclination S0--Sd galaxies.  They conclude that galaxy color 
gradients are due in most part to gradients in age and metallicity in their 
stellar populations (in the sense that inner regions are older and more metal 
rich than outer regions) and contend that dust reddening mainly affects the
metallicity gradients, but is likely too small to affect their conclusions.
%We conclude that dust reddening will mainly affect the metallicity 
%(and to a lesser extent age) gradients: however, the magnitude of dust 
%effects is likely to be too small to significantly affect our conclusions.
In comparing the SFHs with galaxy parameters, they conclude that the SFH 
of a galaxy is primarily driven by surface density and that the total stellar
mass of a galaxy is a less important parameter that correlates significantly 
with metallicity, but not age.  Kauffmann \etal\ (2003a) 
also conclude that the recent SFHs, as probed by the H$_{\delta A}$ 
absorption line index and the 4000\AA\ break, 
%strength age indicators, 
of over 100,000 galaxies are more strongly correlated with 
surface mass density than stellar mass.  BdJ00 argued that these correlations 
could be the result of a surface density-dependent star formation law, 
coupled with galaxy mass-dependent chemically enriched gas outflows.
Bell \& Bower (2000) further explored this idea by constructing a family of 
simple models for spiral galaxy evolution for comparison with the 
observational trends in SFH with galaxy parameters. 
Indeed they found that the data are consistent with the proposition that the 
SFH of a region within a galaxy depends primarily on the local surface 
density of the gas but that additional ingredients, such as galaxy mass 
dependent infall(outflow) of primordial(metal-enriched) gas and/or formation 
epoch, are required to fully explain the observational results.

In spite of genuine progress in recent years in the study of 
spectro-photometric gradients in spiral galaxies, our ability to model and
interpret them in terms of formation models is still limited due to the lack
of data and the intricacies involved in converting observed quantities to 
reliable ages and metallicities.  In this study, we use the largest catalog of 
deep optical and NIR galaxy colors to date to revisit the comparison of 
broad-band color gradients with stellar population models using a range of 
SFHs and basic assumptions about the dust distribution.  We follow the 
approach developed by BdJ00, exploring additional parameter ranges and using 
an extended data base by combining the BdJ00 data with our own collection of 
deep, optical and NIR surface brightness profiles 
(Courteau, Holtzman, \& MacArthur 2004, in prep.; hereafter Paper III), 
for a total sample of 172 galaxies.  For the combined data base we determine 
local average ages and metallicities 
in radial bins for each galaxy and compute gradients in age and metallicity 
as a function of radius.  We pay special attention to the effects of fitting 
out to different physical extents for individual galaxies and the distinction 
between inner and outer galaxy gradients.  In particular, we find that false 
trends can be inferred if the radial extent of the gradient fits is not taken 
into account, which in turn leads to erroneous conclusions about the galaxy
parameters that drive their SFHs.

The outline of the paper is as follows.  In \se{data}, we describe the data 
base from which color gradients are computed.  Our color gradients are 
presented in Paper III, and further details of the BdJ00 sample can be found 
in their \S2.  In \se{params} we explore the range of galaxy parameters in 
our sample and their intrinsic correlations which must be considered when 
comparing trends in age and metallicity gradients with galaxy parameters. In
\se{SSPs}, we discuss the stellar population models to be compared 
to the data and the different star formation prescriptions 
adopted.  Dust models and its potential effects on our results are 
discussed in \se{dust}.   Optical-NIR color-color profiles and their matching
population models are presented in \se{profiles} and trends with galaxy 
parameters are explored.  The technique by which the data are fitted to the 
SSP models is presented in \se{fits}.  Local and global age and metallicities 
are presented in \se{global} and their gradients are discussed in \se{age} 
and \se{metal} and contrasted with previous results from BdJ00. 
We conclude with a discussion of the mechanisms that control stellar 
evolution in spiral galaxies and compare our results with existing
models of galaxy evolution in \se{discussion}. 

\section{The Data}\label{sec:data}

Our data base from which color gradients are measured is a combination of the
data in BdJ00 and the compilation by Courteau et al.(2004, in prep., 
Paper III) of 1063 digital images in the $B$, $V$, $R$, and $H$ passbands of 
324 nearby late-type spiral galaxies collected at the Lowell Observatory and 
Kitt Peak National Observatory.  For the current analysis we 
consider only the face-on and moderately tilted galaxies ($i <$ 60$^\circ$) in 
the sample.  All galaxies were selected from the Uppsala General Catalogue of 
Galaxies (UGC, Nilson 1973) to have: Zwicky magnitude $m_B \leq$ 15.5, blue 
Galactic extinction $A_B=4 \times E(\bv) \leq$ 0\farcm5 \citep{BursHeil84}, 
and blue major axis $\leq 2\farcm2$.  For the computation of homogeneous 
structural parameters and color gradients we use the isophotal map determined 
at $R$-band and applied onto all other images ($BVH$) of a galaxy.  The SB 
profiles are reliable (with SB errors $<$ 0.12 mag arcsec$^{-2}$) down to 
$\sim$ 26 mag arcsec$^{-2}$ for optical bands and $\sim$ 22 mag arcsec$^{-2}$ 
at $H$-band.  For 
the purpose of this color-based analysis, we further require that the galaxies 
have measured surface brightnesses in at least two optical and one near-IR 
($H$) band observations out to at least 1.5 $H$-band disk scale lengths and 
with SB errors of less than 0.12 mag arcsec$^{-2}$.  This leaves us with
51 galaxies (25 type I, 17 type II, 9 transition)\footnote{For type I SB 
profiles (Freeman 1970), the inner profile always lies above the SB of the 
inward extrapolation of the outer disk, whereas type II systems have a portion 
of their brightness profiles lying below the inward disk extrapolation.  We 
define a transition case for luminosity profiles that change from type II at 
optical wavelengths to type I in the infrared.  See Paper I for
a definition of the different profile types.} with extended reliable color 
gradients.  

Radial profiles were corrected for Galactic extinction using the reddening
values of Schlegel \etal\ (1998).  We do not attempt to correct our SB profiles
for internal extinction, but discuss the possible effects of dust on our 
results in \se{dust}.  The SB profiles were degraded to the worst seeing FWHM 
before computing color gradients.  For a full description of the data see 
Paper III.

In order to increase the signal-to-noise (S/N) of our color profiles, the SB
profiles were averaged into radial bins scaled by the NIR-band disk scale 
length, $h_{{\rm NIR}}$, and we required at least 3 data points per bin.  The
bin sizes were defined as follows:

$\quad \quad \qquad grid_1:  \quad \quad \quad \qquad \qquad grid_2: $
\begin{equation}
\begin{array}{cccccccccc}
        0.0 & \le & r/h & < & 0.5, & \quad 
        0.0 & \le & r/h & < & 0.25,  \\ 
        0.5 & \le & r/h & < & 1.5, & \quad      
        0.25 & \le & r/h & < & 0.5,  \\ 
        1.5 & \le & r/h & < & 2.5, & \quad      
        0.5 & \le & r/h & < & 0.75, \\
        2.5 & \le & r/h & < & 3.5, & \quad      
        0.75 & \le & r/h & < & 1.0, \\
        3.5 & \le & r/h & < & 4.5, & \quad      
        1.0 & \le & r/h & < & 1.5,  \\ 
        4.5 & \le & r/h & < & 5.5 & \quad       
        1.5 & \le & r/h & < & 2.0, \\ 
          & &  & &   &\quad       
          2.0 & \le & r/h & < & 3.0
\end{array}
\label{eq:binning}
\end{equation}

The second, finer, grid ($grid_2$) was used to see if the coarser binning
of $grid_1$ hides any important features in the color profiles.  The 
difference in using the two different binnings is small.  In particular, this
test confirmed that the measured gradients do not depend on the central 
pixels, which are likely to be more affected by a central concentration of
dust in spiral galaxies.  Hence we subsequently use $grid_1$ only (as this is 
the binning of the BdJ00 sample).

In order to look for trends in color gradients with galaxy parameters, we 
collected, for as many galaxies as possible, their morphological type, absolute 
magnitude, $H$-band disk central surface brightness (CSB), $H$-band disk scale 
length, inclination, total $H$-band magnitude, and rotational velocity 
$V_{\mathrm{rot}}$ (see below).  The structural parameters (disk scale length, 
$h$, and CSB, $\mu_0$) were computed using our bulge-to-disk decomposition 
scheme (see Paper I).  Total galaxy magnitudes encompass the light out to SB 
levels given above, and include an
%are measured 
%non-parametrically out to the extent of the SB profile (Paper III).
%An extrapolated magnitude is computed by 
extrapolation of the SB profile to 
infinity with an exponential fit to the last quarter of the SB 
profile\footnote{This accounts for outer disk truncations that would not be 
fit properly using the entire profile from our B/D decompositions in 
Paper I.}.  The extrapolated magnitude increment (0.07 $H$-mag on average) is 
added to the isophotal magnitude which is corrected for Galactic extinction 
(Schlegel \etal\ 1998).  Stellar masses are computed using the prescription 
of Bell \& de~Jong (2001).

We merged our sample with the 121 galaxies from BdJ00 which spans a wider 
range in Hubble type (S0--Irr) and includes
a few low surface brightness (LSB) galaxies.  For the computation of color
gradients, BdJ00 required at least two optical ($BVRI$) and one $K$-band 
observations per galaxy.  The data reductions and radial binning (using 
only $grid_1$) are very similar to our own.  
For 3 galaxies in common with both data sets, there is excellent 
agreement in the overall calibration and radial profiles at least for 
the $BVRH$ bands (Paper III).

In order to compare the two data sets directly when looking for trends in 
age and metallicity with galaxy parameters, we converted our $H$-band
magnitudes and surface brightnesses into $K$-band using the 2MASS $H-K$ color
transformations derived from Jarrett \etal\ (2003).  
\begin{deluxetable}{cccccccc}
\tabletypesize{\normalsize}
%\tablewidth{7cm}
\tablecaption{\label{tab:2mass}}
\tablehead{
\colhead{Hubble Type} &
\colhead{Sa} &
\colhead{Sab} &
\colhead{Sb} &
\colhead{Sbc} &
\colhead{Sc} &
\colhead{Scd} &
\colhead{Sd} }
\startdata
%Hubble Type & Sa & Sab & Sb & Sbc & Sc & Scd & Sd \\
mean $H-K$  & 0.30 & 0.26 & 0.30 & 0.28 & 0.28 & 0.24 & 0.21 \\
\tablecomments{Mean 2MASS $H-K$ isophotal colors for non-barred galaxies. 
              Adapted from Figure~16 of Jarrett \etal\ (2003).}
\enddata
\end{deluxetable}
Since we are interested 
mainly in qualitative trends in age and metallicity with galaxy parameters, 
coarse transformations are adequate and we took the mean $H-K$ values for 
each Hubble type, derived from their Figure~16 and listed here in 
Table~\ref{tab:2mass}.  The modified $H$ magnitude will be denoted as 
$H_{\mathrm{m}}$, to represent the $K$-band equivalent of the $H$-band 
measurement.  We do, however, consider the disk scale lengths in the $H$ and 
$K$ to be directly comparable.  M{\" o}llenhoff \& Heidt (2001) find 
$h_J/h_K = 1.01 \pm 0.19$ for a sample of 40 spiral galaxies and the 
relationship will be even tighter for $h_H/h_K$, hence a direct comparison
of the $H$ and $K$ band disk scale lengths is justified.

It has been suggested that age and metallicity gradients may depend upon the 
potential of the galaxy and thus vary as a function of $V_{\mathrm{rot}}$
(Dalcanton \& Bernstein 2002; Garnett 2002a).  In order to search for such a 
trend we augmented out catalog with \ion{H}{1} or H$\alpha$ line widths 
(which we take to be $W50/2$, half the line width at 50\% peak flux) 
collected from the literature.  These data will enable us to test for a 
distinct signature in the color profiles from the more prominent dust lanes 
seen in galaxies with 
$V_{\mathrm{rot}} >$ 120 \kms\ (Dalcanton \& Bernstein 2002) and/or if 
there is a flattening of the metallicity of galaxies above 
$V_{\mathrm{rot}} \sim$ 150 \kms\ (Garnett 2002a).  The 
line widths were collected from various sources  
%(Bottinelli \etal\ 1990 [74 gals]; Theureau \etal\ 1998 [11 gals]; 
%Courteau 1997 [7 gals]; de~Blok, McGaugh, \& Rubin 2001 [4 gals]; 
%Haynes \etal\ 1999 [1 gal]; de~Blok \& Bosma 2002 [1 gal])
and transformations were derived to ensure uniformity between them 
(see Appendix~\ref{HIwidths}).  Differences between 
measurements for a given galaxy for different sources were typically less 
than 10\%, which is sufficient for our purposes.  

In order to ensure sufficiently accurate $V_{\mathrm{rot}}$ measurements, we 
restricted our sample to galaxies with inclination greater than 35$\degr$. 
We are left with 98 galaxies in our sample that have reliable rotational 
velocity measurements.

\section{Galaxy parameters}\label{sec:params}

Trends in age and metallicity (local values and gradients) with galaxy 
parameters, may bear an imprint from intrinsic correlations within the sample 
itself.  In Figure~\ref{fig:galpars} we plot the galaxy parameters used in 
this analysis ($h$, $\mu_0$, $M_{H_\mathrm{m},K}$, and $V_{\mathrm{rot}}$) 
against each other.  The most notable correlations are the 
luminosity-rotation speed relation, $M_{H_\mathrm{m},K}$ versus 
log($V_{\mathrm{rot}}$) (the so-called ``Tully-Fisher'' relation), and the 
size-luminosity relation, $M_{H_\mathrm{m},K}$ versus log($h$).
All correlations with $\mu_0$ are weaker and have large scatter.  
In Figure~\ref{fig:pars_morph} we plot the 4 basic galaxy parameters as a 
function of morphological type.  
Trends are seen with galaxy parameters for 
$T \ga 4$ such that later-type galaxies 
have smaller rotational velocities and total magnitudes, and lower central 
surface brightness.  On the other hand, there is no clear trend of disk 
scale length with Hubble type (and hence the trend with total magnitude is 
largely driven by the trend with CSB, Courteau \etal\ 2003). Data points are 
few for $T < 4$, but deviations from the $T \ga 4$ trends seem to occur.  A 
larger spread is (barely) noticeable in $V_{\mathrm{rot}}$, but is further 
supported by the larger spread in $M_{H_\mathrm{m},K}$ for which we have more 
galaxies (unfortunately, reliable $V_{\mathrm{rot}}$s could not be found for 
the earliest-type galaxies.)  The increasing trend in $\mu_0$ with decreasing 
$T$ seems to level off at $T < 4$, but the spread is still significant.  
Finally, earliest-type galaxies have the shortest scale lengths, as 
expected.  We shall return to these correlations in the analysis of trends 
in age and metallicity gradients (\se{results}).

\section{Stellar Population Models} \label{sec:SSPs}

In order to translate color information into constraints on the underlying 
ages and abundances of the stellar populations, the color gradients must 
be compared with stellar population synthesis models.  In their most basic 
form, commonly referred to as simple stellar populations (SSPs), these models 
provide evolutionary information for a coeval population of stars born with 
a given composition and initial mass function (IMF).  The closest physical
analog to such SSPs are globular cluster systems from which the SSP models 
are calibrated (\eg\ Schiavon \etal\ 2002).  Several such SSP models have 
been produced by a number of independent groups and are in a constant state 
of modification as improvements to many of the input parameters 
(\eg\ stellar libraries, model atmospheres, convection, mass loss, mixing) 
come to light.  There are discrepancies among the different models that, 
depending on the application, may result in significantly different 
interpretations of observations of unresolved stellar populations.  In order 
to determine if these model discrepancies could affect our analysis, we have 
compared two independent sets of SSPs; the 2003 implementation of the 
Bruzual \& Charlot (2003) models (hereafter referred to as GALAXEV), and the 
Project d'\'{E}tude des Galaxies par Synth\`{e}se \'{E}volutive (PEGASE 2.0) 
models of Fioc \& Rocca-Volmerange (1997).

The stellar populations of interest here are those of spiral galaxies spanning 
the full Hubble sequence (S0--Irr).  The approximation of a single evolved 
stellar population clearly does not apply to evolved, complex stellar 
populations of spiral galaxies.  However, if one assumes that
galaxies are composed of a superposition of SSPs, born at different epochs, 
rates, and metallicities, one can use the SSPs to develop model grids that 
mimic the range of plausible galactic stellar populations and these can be
compared with galaxy color profiles. The stellar population model grids are 
created by taking the single burst SSPs, with constant stellar IMF and fixed 
metallicity, and convolving them with a given SFH.  A few simplifying 
assumptions are inherent in this formulation.  We are assuming that the IMF 
does not change as a function of time or galactic environment.  The validity 
of this assumption may be questioned from a theoretical perspective as the 
IMF is expected to vary systematically with star formation environments 
(Chabrier 2003, and references therein).  However, significant observational 
evidence of star formation in small molecular clouds, rich and dense massive 
star-clusters, as well as ancient metal-poor stellar populations, reveals a 
remarkable uniformity of the IMF (Kroupa 2002).  The use of fixed-metallicity 
SSP implies no chemical evolution.
This is clearly an unrealistic assumption with regards to galaxy evolution
(as it also ignores feedback effects from stellar winds and supernovae), but
it still allows us a reasonable comparison of {\it relative} metallicities and
ages within and among galaxies (Abraham \etal\ 1999; BdJ00; 
Gavazzi \etal\ 2002).

The specific choice of IMF, parameterized as 
$\xi(\log m) \propto m^{-x}$, can 
also potentially affect our results.  To gauge the importance of this effect 
on our color-based analysis we compared model grids obtained with some of the 
most widely used IMF characterizations; (a) the single-slope Salpeter (1955) 
with 

\begin{eqnarray}
x & = & 1.35,\qquad 0.1 \le m/M_{\odot} \le 100,
\label{eq:salp}
\end{eqnarray}

(b) the Kroupa (2001) IMF (PEGASE models only) with:

\begin{eqnarray}
x_{0} & = & 0.3,\qquad 0.1 \le m/M_{\odot} \le 0.5\,\,\,
\nonumber \\
x_{1} & = & 1.3,\qquad 0.5 \le m/M_{\odot} \le 120,
\label{eq:kroupa01}
\end{eqnarray}

and (c) the Chabrier (2003) single-star Galactic disk IMF 
(GALAXEV models only) which is parameterized as: 

\begin{equation}
\xi(\log m) \propto \cases{\exp\left[-\frac{(\log m - \log m_{c})^{2}}{2\sigma^2}\right],& $0.1 \le m/M_{\odot} \le 1 $;\cr m^{-1.3},& $1 < m/M_{\odot} \le 100 $}
\label{eq:chab}
\end{equation}

The Chabrier (2003) IMF is preferred over the Kroupa (2001) and Salpeter 
(1955) IMFs because of its better agreement with number counts of brown dwarfs 
in the galactic disk and its theoretical motivation.

The integrated spectrum, $F_{\lambda}(t)$,  of a stellar population with SFR 
$\Psi(t)$ is computed as the convolution of a single
burst stellar population of given metallicity, $f_{\lambda}(t)$, with the
given SFR:

\begin{equation}
F_{\lambda}(t) = \int^{t}_{0}\Psi(t-t')f_{\lambda}(t')\,\mathrm{d}t'.
\label{eq:intspec}
\end{equation}

The left panel in Figure~\ref{fig:tracks_imf} compares the model tracks from 
the GALAXEV models with the Salpeter (1955) (eq.~\ref{eq:salp}) and Chabrier 
(2003) (eq.~\ref{eq:chab}) IMFs, and the right panel compares those from the 
PEGASE models for Salpeter (1955) and Kroupa (2001) (eq.~\ref{eq:kroupa01}) 
IMFs.  The difference between the various IMFs on the model grids are 
small, thus the specific choice of IMF (within current observational 
constraints) will not affect our results.  We opt to use the Salpeter (1955) 
IMF for the remainder of this analysis (primarily to facilitate comparison
with other studies).  

Also over-plotted in the right panel of Figure~\ref{fig:tracks_imf} are the 
GALAXEV model with Salpeter IMF tracks for direct
comparison with the PEGASE model tracks for the same IMF (although note that
the upper mass cut-off is 100~M$_\odot$ in the GALAXEV models and 
125~M$_\odot$ in the PEGASE models).  While differences in the grids between 
the different models are noticeable (most likely due
to different treatments of the thermally pulsing asymptotic giant branch 
phase, see e.g. Maraston 2003), they are 
only significant in the bluest regions where the SB errors of our observations
are also large (\ie\ the outer regions of our spiral galaxies where sky 
subtraction errors become large, are, for the most part, bluer).  Slight 
differences in the age and metallicity gradients would be inferred using the 
different models, but we are mainly interested in relative quantities, and 
thus are not concerned with the small absolute differences between models. 
For the rest of this analysis we refer only to the GALAXEV models.

We have explored two different SFR regimes, both parameterized by a star 
formation timescale $\tau$.  The first is the simple exponential SFR, 
$\Psi_{\mathrm{exp}}(t)$,

\begin{equation}
\Psi_{\mathrm{exp}}(t) = \frac{c_{\mathrm{exp}}}{\tau} e^{-t/\tau},
\label{eq:exp}
\end{equation}

where

\begin{equation}
c_{\mathrm{exp}} = \frac{1}{1 - e^{-A/\tau}}.
\end{equation}

The second is the so-called ``Sandage'' SFR, $\Psi_{\mathrm{San}}(t)$ 
first parameterized by Sandage (1985),

\begin{equation}
\Psi_{\mathrm{San}}(t) = c_{\mathrm{San}} \frac{t}{\tau^{2}} e^{-t^2/2\tau^2}
\label{eq:sandage}
\end{equation}

and

\begin{equation}
c_{\mathrm{San}} = \frac{1}{1 - e^{-1/2(A/\tau)^{2}}}
\end{equation}

where $c_{\mathrm{exp}}$ and $c_{\mathrm{San}}$ are normalization constants 
(to 1 M$_{\odot}$).  Figure~\ref{fig:sfr_time} shows the time evolution of the 
exponential (top panel) and Sandage SFRs (bottom panel).  
The exponential SFR 
falls off at a given rate for all positive values of $\tau$.  However, in order
to cover the full color-color space spanned by the observations, negative 
values of $\tau$ were also included which allow for increasing SFRs.  The 
Sandage SFR is characterized by a delayed rise in the SFR, followed by
an exponential decline, the rate of which is determined by the value of $\tau$.
For values of $\tau > A$, where $A$ is the age of the galaxy (fixed to
13 Gyr in this work), the SFR is still rising at the present time.  The 
average 
age of a stellar population of a given $\tau$ is computed as:

\begin{equation}
\langle A \rangle = A - \frac{\int_{0}^{A} t \Psi(t) \,\mathrm{d} t}
                                   {\int_{0}^{A} \Psi(t) \,\mathrm{d} t}.
\label{eq:avgage}
\end{equation}

For the exponential SFR, this equates to,

\begin{equation}
\langle A \rangle_{\mathrm{exp}} = 
             A - \tau \frac{1-e^{-A/\tau}(1+A/\tau)}{1-e^{-A/\tau}}
\label{eq:avgageexp}
\end{equation}

and for the Sandage SFR,

\begin{equation}
\langle A \rangle_{\mathrm{San}} = 
             A - \frac{-Ae^{-\frac{1}{2}\frac{A^2}{\tau^2}} + \frac{1}{2}\tau\sqrt{2\pi}\,\mathrm{erf}\left(\frac{A}{\sqrt{2}\tau}\right)}{1-e^{-\frac{1}{2}\frac{A^2}{\tau^2}}}.
\end{equation}

Figure~\ref{fig:tracks_sfh} shows the resulting model grids for the exponential
(blue), and Sandage (red) SFRs overlaid on each other.   Lines of constant 
average age $\langle A \rangle$ are the dashed, 
roughly vertical lines (labeled in Gyr), while lines of constant metallicity 
are solid and run closer to the horizontal (labeled in Z, the mass fraction in
elements heavier than helium, where Z$_{\odot} = 0.02$).  
The overall shape
of the two model grids are quite similar, the main difference being that
the Sandage SFH has a younger average age for a given $B-R$ and
the metallicity is slightly higher for a given $R-H$.  The Sandage grids
also cover a wider range in ages (which would increase the magnitude of our 
measured age gradients).

The reported radial age gradients inferred from color gradients 
(\eg\ BdJ00) seem to be at odds with the lack of age gradient observed in 
open clusters of the Galaxy (Freeman \& Bland-Hawthorn 2002).  Would it be 
possible that the apparent radial age gradients observed in spiral galaxies 
are simply due to 
small amounts of young frostings of star formation on top of an 
underlying old population?  In order to test for such effects in our age 
determinations, we produced new model grids for the same underlying SFH but 
with an additional burst of star formation at different times and fractional
masses.  The burst was added as a Sandage (1985) type profile 
(eq.~\ref{eq:sandage}) with $\tau = 0.1$\footnote{This timescale is justified 
on the basis that starbursts are consistent with a constant SF lasting 
10--100 Myr (Meurer 2000).}.  The average age for the models with a burst 
superimposed on the underlying SFR is now computed as,

%\begin{equation}
%\langle A \rangle = A - 
%       \frac{(1-b_{frac})\int_{0}^{A} t \Psi(t) \,\mathrm{d} t + b_{frac}\int_{t_{burst}}^{A} (t-t_{burst})\Psi_{burst}(t-t_{burst}) \,\mathrm{d} t }
% {(1-b_{frac})\int_{0}^{A} \Psi(t) \,\mathrm{d} t  + 
% b_{frac}\int_{t_{burst}}^{A} \Psi_{burst}(t-t_{burst}) 
% \,\mathrm{d} t }
%\label{eq:avgageburst}
%\end{equation}

\begin{equation}
\langle A \rangle = A - 
       \frac{(1-b_{f}) \int_{0}^{A} t \Psi(t) \,\mathrm{d} t + b_{f}\int_{t_{b}}^{A} t'\Psi_{b}(t') \,\mathrm{d} t }
 { (1-b_{f})\int_{0}^{A} \Psi(t) \,\mathrm{d} t  + 
 b_{f}\int_{t_{b}}^{A} \Psi_{b}(t') 
 \,\mathrm{d} t }
\label{eq:avgageburst}
\end{equation}

where $b_{f}$ is the mass fraction of the burst, $t_{b}$ is the time 
since the burst, \ie\ the burst of star formation occurs $t_{b}$ Gyr 
before the current age of the galaxy, $A$, and $t' = t - t_{b}$.  
Figures~\ref{fig:tracks_burst1} and \ref{fig:tracks_burst2} demonstrate the 
effect of such a burst of star formation superimposed on an underlying 
exponential SFR with ages ($t_{b}$) of 1 and 4 Gyr respectively.  The left 
panel in both figures shows the effect if 10\% of the galaxy's total mass 
was involved in the burst, and the right panel shows a 50\% (by mass) burst.  
The models certainly do not cover the same extent in the color-color space 
for recent bursts.
With only 10\% of the total galaxy mass in the burst 
(Fig.~\ref{fig:tracks_burst1}; left panel) the grid does not extend to the red 
as far as those without bursts.  The effect is strongest ($\sim$ 0.3 mag) 
in the optical colors.  Also, the burst grid does not extend as far into 
the blue optical colors ($\la$ 0.1 mag) for the most metal-rich galaxies.  This
is because we are essentially adding more older stars to the SFHs that are
rising at 13 Gyr ($\tau < 0 $ for $\Psi_{exp}$, or $\tau > 13$ Gyr for 
$\Psi_{San}$; This effect is stronger at higher metallicity because the
evolution of $B-R$ color with time steepens for higher metallicity
stellar populations at ages $\sim$ 1--3 Gyr in the GALAXEV models.)  The
lines of constant age also steepen.  ``Frosting'' also shifts the $R-H$ colors 
to redder values ($\sim$0.1 mag) at a given metallicity.  This, in fact, 
prevents the burst grid from reaching red-ward enough as required by the 
observational data for the inner-most parts of the galaxies 
(see Figs.~\ref{fig:grads_morph}--\ref{fig:grads_sb}).  Either the bursts are 
accompanied by significant amounts of dust, or 1 Gyr old bursts can be ruled 
out for the central parts of late-type galaxies.  The 50\% by mass burst in
Figure~\ref{fig:tracks_burst1}, right panel, shows that the grid covers only
a narrow range in color--color space that does not extend far enough into
the red or the blue to agree with the observations.  For older bursts 
of star formation, the model grids have colors fully consistent with the 
no-burst models.  This is shown in Figure~\ref{fig:tracks_burst2} for a pure 
exponential SFH (blue) overlaid with an exponential plus $t_{burst} =$ 4 Gyr 
Sandage-like burst (red).  The grids are almost identical with respect to 
their shape and location (with slight narrowing of the 50\% mass burst grid).  
The only difference is the average ages (eqs.~\ref{eq:avgage} \& 
\ref{eq:avgageburst}), being younger for a given color of the burst grid.  
Thus, adding a burst of star formation that is older than $\sim 1$ Gyr has 
a similar effect as changing the overall form of the SFH; the grid shape and 
color coverage are the same, but the average age can change significantly 
from one scenario to another.
Not only do the age determinations change, but the age gradients can also be
different for a given SFH (compare Sandage and exponential SFH model grids in
Fig.~\ref{fig:tracks_sfh}).  Since the actual SFHs of real galaxies are 
unknown, the {\it absolute} values of the gradients measured with this 
color-based technique cannot be trusted.  We can, however, trust that a 
gradient exists in a galaxy, but not its magnitude.  We can also compare
relative gradients among different galaxies if we assume self-similar SFHs 
and no significant bursts of star formation within the last 1--2 Gyr.

A reasonable estimate of the validity of this assumption can be obtained from
the study of Kauffmann \etal\ (2003a) who used the H$\delta_{A}$ absorption 
line index and the 4000\AA\ break age indicators along with the 
Bruzual \& Charlot SSPs to constrain the SFHs, dust attenuation, and 
stellar masses of over 100,000 galaxies from the Sloan Digital Sky Survey 
(SDSS).  For each galaxy they estimated F$_{burst}$, the fraction of the 
total stellar mass formed in bursts in the past 1-2 Gyr.  Their Figure~5 
shows the fraction of galaxies with F$_{burst}>0$ at the 50\% and 97.5\% 
confidence levels in stellar mass bins ranging from 
8.0--12.0~log$_{10}(M_\odot)$.  From this and estimates of the galaxy stellar 
masses (Bell \& de~Jong 2001), we infer the fraction of 
galaxies in our sample that have undergone bursts of SF in the past 1--2 Gyr 
is $\sim 1$\% ($\sim 12$\%) at the 97.5\% (50\%) confidence level.  
The fraction of galaxies that have F$_{burst} \gg{10}$\% will be even smaller
and we conclude that our assumption of no recent burst is valid.

\section{Dust effects} \label{sec:dust}

While the presence of dust in late-type spiral galaxies is well-established,
its distribution  
and effective optical depth remain poorly constrained.  Extinction by dust 
could potentially mimic a color gradient and its effects must be 
considered in determining the stellar content of galaxies from 
colors.  We attempt to quantify the effects of dust extinction on colors with
simple dust geometry extinction models,  adopting the MW extinction curve and 
albedo values from Gordon, Calzetti, \& Witt (1997).  The extinction vector 
for the simplistic foreground screen model with $A_V = 0.3$ is shown (as the 
arrow) in the upper left corner in 
Figures~\ref{fig:grads_morph}--\ref{fig:grads_sb}.  The
vector is in the same general direction as the observed gradients of the 
galaxies, but given the unrealistic nature of this model, we do not consider 
it further (but include it for visual comparison).  Also shown in the top
left corner in Figures~\ref{fig:grads_morph}--\ref{fig:grads_sb} is the more
realistic ``face-on triplex'' dust model of 
Disney, Davies, \& Phillipps (1989) and Evans (1994).  This model assumes
that the stars and dust have exponential distributions in both the radial and 
vertical directions with radial scale lengths $h_*$ and $h_d$,
and vertical scale lengths $z_*$ and $z_d$.  Although scattering is not taken 
into account,  Byun, Freeman, \& Kylafis (1994) and de~Jong (1996) have shown 
that its effects are negligible in face-on galaxies (as photons are just as 
likely to be scattered into the line of sight as they are out of it.)  We 
therefore use the dust absorption curve to compute the dust optical depth;

\begin{equation}
\tau_{\lambda}(0) = \frac{\tau_{\lambda}}{\tau_{V}} \tau_{V}(0) (1 - a_{\lambda})
\label{eq:tau}
\end{equation}

where $\tau_{\lambda}$ and $a_{\lambda}$ are the dust optical depth and albedo
respectively.

Studies of multi-wavelength data that solve simultaneously for the intrinsic 
color of stellar populations and reddening and extinction effects from dust 
find central optical depths, $\tau_{V}(0)$,
in the range 0.3--2.5 for types Sab--Sc (Peletier \etal\ 1995; Kuchinski 
\etal\ 1998; Xilouris \etal\ 1999).  The HST and ground-based color gradient 
study of early-type bulges ($\le$Sb)
by Peletier \etal\ (1999) reveals dust extinctions of A$_V =$ 0.6--1.0 mag
(or A$_H =$ 0.1--0.2 mag) in galaxy centers but negligible extinction beyond 
one effective radius.  Multi-band SB profile modeling of massive 
edge-on disks suggests that the dust is confined to a thin extended plane, 
with $z_{d}/z_{*} \la 0.7$ and  
$h_{d}/h_{*} \sim 1.4$ (Xilouris \etal\ 1999; Matthews \& Wood 2001).
Masters, Giovanelli, \& Haynes (2003) studied internal extinction 
in the near-IR for a sample of 15,244 2MASS galaxies by looking at the
inclination dependence of various photometric parameters, and 
concluded that galaxies with M$_{\lambda} - 5$log(h) $>$ 
-20, -20.7, and -20.9 in the  $J$, $H$, and $K_s$-bands respectively are
spared any extinction, but that disk opacity increases 
monotonically with disk luminosity.  Application of the triplex model for 
these galaxies favors small dust-to-star scale 
height ratios ($z_{d}/z_{*} \sim 0.5$, in good agreement with Xilouris \etal\ 
1999) and face-on central 
opacities of $\tau\degr(0) = 0.7, 0.3$ at $H$ and $K_s$ respectively.

We adopt the same vertical and radial scale length ratios for the dust and
stars; using $z_{d}/z_{*} = 0.7$ and  $h_{d}/h_{*} = 1.4$ instead stretches 
the gradients by only 0.018 in $B-R$, 0.019 in $R-H$, and 0.021 in $R-K$, 
and the central $V$-band optical depth to $\tau_{V}(0) = 1$ (or pole-to-pole 
$V$-band optical depth of 2).  The higher end of the measured range of 
$\tau_{V}(0) = 2$ is shown in the upper left corner of 
Figures~\ref{fig:grads_Vrot} \& \ref{fig:gradsK_Vrot} (right panels).

The triplex models lie parallel to, and thus could contribute significantly 
to, the observed gradients (note there is no overall calibration in the dust 
models so they can slide to any region on the plot).  For the gradients to be 
entirely due to dust requires extremely high central optical depths 
($\tau_{V}(0) \sim 5$) to reproduce the data.  Also, the triplex models alone 
cannot reproduce the significant color gradients observed from the half-light 
radius (denoted by open circles) outward.  Thus, while dust is likely a
contributor to color gradients, age and metallicity effects must also be
invoked.

Dalcanton \& Bernstein (2002) have analysed optical and infrared color maps of 
47 extremely late-type (bulge-less) edge-on spirals spanning a wide mass range 
(40 $<$ V$_c <$ 250 \kms).  They find that higher mass (rotation velocity) 
galaxies have more prominent dust lanes and have redder colors than the 
stellar population grids (whereas the less massive galaxies have colors 
consistent with the population grids).  Analysis of their $R-K_s$ color maps, 
and a comparison of colors between their edge-on sample and the face-on 
sample of de~Jong (1996) suggests that dust plays little role in all but the 
most massive galaxies in their sample (V$_c >$ 120 \kms).

Previous analyses have ruled out dust effects on the color gradients in 
galaxies based on the fact that they see no correlation between the 
gradients and inclination.  Dust effects on gradient profiles as a function 
of bulge-to-total ratio (B/T) and inclination were modeled by 
Byun, Freeman, \& Kylafis (1994).  Their Figure~6 shows $B-I$ versus radius 
for model galaxies with different B/T ratios and $\tau_{V}(0)$ ranging from 
0.0--10.0 for inclinations 0--85\degr.  In the 0--50\degr\ range the 
differences in the profiles are very small and would not be detected in the
observations.  Our measured radial gradients also show no inclination 
dependence, but we do not consider this grounds to rule out significant dust 
extinction.

Based on optical-IR imaging of S0--Sbc galaxies, Peletier \& Balcells (1996) 
found that ``dust-free'' colors of galaxy disks are not significantly 
different from their bulges.  They derived bulge colors from minor-axis wedges 
in images of early-type edge-on spirals.  The wedges are presumed dust-free 
above the disk plane.  They conclude that (dust-free) bulge and disk colors 
are very similar with $\Delta(B-R) = 0.045 \pm 0.097$ and 
$\Delta(R-K) = 0.078 \pm 0.165$.  We find a significantly different result;
as can be seen in Figure~\ref{fig:bulge_disk} (compare with Fig.~2 in 
Peletier \& Balcells 1996), our bulges are much redder than their disks with
$\Delta(B-R) = 0.29 \pm 0.17$ as $\Delta(R-K) = 0.30 \pm 0.17$.  
This could be 
due to genuine dust extinction in our galaxies or there is a fundamental 
difference between our respective samples and/or analysis methods.
Their disk SB profiles were measured along 10\degr-wide wedge 
apertures centered 15\degr\ away from the disk major axis, to avoid the
prominent dust lanes near the major axis of their inclined galaxies.  
Naturally, this technique is sensitive to a vertical disk color gradient.  
Dalcanton \& Bernstein (2002) have suggested that all thin disks are 
embedded in a red stellar envelope.  Whether this envelope is redder or bluer
than the thin disk depends on the rotational velocity of the galaxy: 
redder envelopes for $V_{rot} < 120$ \kms\ and bluer otherwise, but the
color of the envelope is similar for all disk galaxies.  The redder thin
disks of the galaxies with $V_{rot} > 120$ \kms\ are attributed to 
strong dust lanes observed in their $B-R$ color maps, which disappear in 
the $V_{rot} < 120$ \kms\ galaxies.  Clearly, these red stellar 
envelopes, and hence the presence of vertical color gradients in disk galaxies 
renders the interpretation of Peletier \& Balcells (1996) ``disk'' colors 
difficult. The 10\degr\ offset from the major axis may result in a measurement 
of the ``red envelope'' stars which are intrinsically older 
than the thin disk stars (though not necessarily redder since thin disk 
stars could be reddened by a central dust lane).

While optical-IR color imaging alone is not sufficient to break the 
degeneracy between dust and stellar population effects on the color 
gradients, the tentative consensus to date is that 
dust is generally not a significant contributor to galaxy colors in 
low-mass/low-luminosity spiral galaxies, but is likely important in more 
massive/brighter galaxies.  It must be borne iin mind that future studies of
high-resolution IR and FIR imaging and absorption-line spectroscopy may 
radically alter this view.

\section{Color-color Profiles}\label{sec:profiles}

Optical-NIR color-color profiles are shown for the Courteau \etal\ sample of 
in Figures~\ref{fig:grads_morph}--\ref{fig:grads_Vrot} and 
\ref{fig:grads_Mh}--\ref{fig:grads_sb} (see Figs.~1--5 in BdJ00 for similar 
plots with their sample and our Fig.~\ref{fig:gradsK_Vrot} which is only
a subset of the BdJ00 sample with available $V_{\mathrm{rot}}$ valuess).
Typical observation errors are shown as crosses in the lower right corners of 
Figures~\ref{fig:grads_morph}--\ref{fig:grads_sb}.  From left to right the
crosses represent calibration errors, and average sky subtraction errors for
the innermost and outermost bins.
The galaxy centers are indicated by solid symbols: circles for type-I galaxies,
triangles for type-II, and asterisks for transition galaxies (as defined in
Paper I).  Open circles denote the half-light radius of the disk.  The line 
types for the galaxy color profiles represent bar strength: solid for bar-less 
(A), dashed for mild bars (AB), and dot-dashed for strong bars (B).  It is 
conceivable that the type-II ``dip'' could be due to dust extinction and/or
stellar population effects, possibly linked to the presence of a bar,
and or inner-disk truncation (see Paper I).  However, we do not see any 
distinction between the 
different SB profile types in the color gradients.  This argues against dust
as a major factor for the type-II phenomenon as dustier galaxies would be 
redder and have larger gradients.  No conclusion can be drawn about the 
effects of a bar due to the small size of our sample and the fact that 
non-barred galaxies may have once harbored a bar that dissolved after mixing
the stellar population.

Over-plotted on Figures~\ref{fig:grads_morph}--\ref{fig:grads_sb} are our 
convolved Bruzual \& Charlot (GALAXEV) stellar population models with an 
exponential SFH
at different metallicities.  Lines of constant average age $\langle A \rangle$ 
(see eq.~\ref{eq:avgage}) are the dashed, roughly vertical lines, while lines 
of constant metallicity are solid and nearly horizontal.  In the 
upper left corner, triplex and foreground screen dust models are 
plotted with the galaxy center and disk half-light radius denoted by solid 
and open circles respectively.  All of the galaxies show significant color 
gradients that are consistent with gradients in both age and metallicity with 
the central parts having older mean ages and being more metal rich.  
Figures~\ref{fig:grads_Vrot} \& \ref{fig:gradsK_Vrot} for the Courteau 
\etal\ and BdJ00 samples reveal a largely consistent picture of the 
radially-resolved colors of spiral galaxies, though the Courteau \etal\ 
galaxies extend to higher metallicity (0.1--0.2 mag in $R-H$).  The origin 
of this discrepancy is not clear, but we note that the optical versus $J-K$ 
colors of galaxies are not well fit by single metallicity models 
(e.g., Fig.~15 of Bell \etal\ 2003).  It may well be that, due to TP-AGB 
prescriptions, systematic errors of 0.2 mag between $H$ and $K$ predicted 
colors are inevitable.  Thermal emission from the telescope at $K$-band may 
also affect sky flats of the type used by BdJ00.
Bearing in mind these potential sources of systematic uncertainty, we feel
that the degree of similarity between the Courteau \etal\ and BdJ00 galaxy
samples is satisfying.

The small number of earlier-type galaxies (Sab--Sbc) in our sample makes it 
difficult to infer any trends in age and metallicity with galaxy type, but
Figure~\ref{fig:dAZ_morph} shows hints that later-types are slightly 
less metal-rich than the earlier types.

In Figures~\ref{fig:grads_Vrot} \& \ref{fig:gradsK_Vrot} we divide the 
Courteau \etal\ and BdJ00 samples, respectively, at $V_{\mathrm{rot}} 
\lessgtr 120$ km$\,$s$^{-1}$, the threshold above which all 
edge-on galaxies have prominent central dust lanes 
(Dalcanton \& Bernstein 2002).  Due to inclination restrictions, we could 
only retrieve reliable \ion{H}{1} line-widths for 28 of our 51 galaxies and 
70 out of the 121 BdJ00 sample.  A trend with $V_{\mathrm{rot}}$ is detected in
Figures~\ref{fig:grads_Vrot} \& \ref{fig:gradsK_Vrot} with faster rotators 
being more metal rich and having older mean ages; also consistent with them 
having a higher dust content.  However, according to the triplex 
models, the true signature of dust is not simply a redder color, but also a 
stretched color profile (compare triplex profiles in 
Figures~\ref{fig:grads_Vrot} \& \ref{fig:gradsK_Vrot} for $V$-band optical 
depths of $\tau_{V}(0) = 1$, left panels, and $\tau_V(0) = 2$, right panels).  
For our sample (Fig.~\ref{fig:grads_Vrot}), the length of the 
gradients appears to be about the same in both velocity bins.  For the BdJ00 
sample (Fig.~\ref{fig:gradsK_Vrot}) the gradients could be more
extended in the larger $V_{\mathrm{rot}}$ bin, but the stretching is not
predominantly along the $R-K$ axis, as in the triplex dust models.  Clearly
we cannot attribute with absolute certainty the redder colors of the
faster rotators to dust effects.

\section{Model Fitting}\label{sec:fits}

Ages and metallicities are determined by fitting the SPS models to the radial
galaxy colors using a maximum-likelihood approach similar to that of
BdJ00 (see their \S3 for further details).  We compute a finely spaced grid 
in $\tau$ using equations~\ref{eq:intspec} \& \ref{eq:exp}, 
and interpolate (linearly) between the SPS metallicities.  Treating the 
colors and errors of each annulus separately, we compute an average 
age and metallicity per annulus by minimizing the \chisqr\ statistic:

\begin{equation}
\chi^{2} = \frac{1}{N-1}\sum_{i=1}^{N}\frac{(\mu_{\mathrm{obs},i}-\mu_{\mathrm{model},i}(\langle A \rangle,Z)-\mu_{c})^2}{\delta\mu^{2}_{\mathrm{tot},i}},
\end{equation}

where $N$ is the number of passbands (at least 3; 2 optical plus 1 IR), 
$\delta\mu_{\mathrm{tot},i}$ is the 
total error in passband $i$, and $\mu_{c}$ is the best normalization
between the model and data computed as a weighted average:

\begin{equation}
\mu_{c} = \frac{\sum_{i=1}^{N}\frac{(\mu_{\mathrm{obs},i}-\mu_{\mathrm{model},i}(\langle A \rangle,Z)-\mu_{c})^2}{\delta\mu^{2}_{\mathrm{tot},i}}}{\sum_{i=1}^{N}\frac{1}{\delta\mu^{2}_{\mathrm{tot},i}}}.
\end{equation}

Errors for the individual age and metallicity measurements as well as their 
gradients are estimated using a Monte Carlo approach.  One hundred 
realizations of the model fits are obtained for 
each radial bin using errors drawn from a normal distribution of the 
observational errors (which include calibration, sky subtraction, and flat 
fielding errors added in quadrature).  For each realization,
gradients are computed as weighted linear fits to the parameter (age and 
metallicity) determinations as a function of the radial scale length.  The 
weights in the fits are taken as the $\Delta \chi^{2} = 1$ interval for each 
annulus--model fit.  The error for the measured gradients and individual 
ages and metallicities are taken as half the interval containing 68\% 
of the 100 Monte Carlo realizations (\ie\ the 1$\sigma$ confidence interval).

A few galaxies in both samples fall outside the model 
grids and thus cannot be fit reliably.  These were removed from the 
sample for the model fits, but are worthy of some discussion.  As can be seen 
in Figure~\ref{fig:grads_morph}, five of our galaxies lie significantly
red-ward of the model grids in their central $R-H$ color.  The
dust vector indicates that such red colors could result from a central dust 
concentration.  Another explanation could be an extremely 
(unrealistically) metal-rich central stellar population.  Three of our galaxies
lie considerably red-ward of the model grids in their $B-R$ colors.  Given
the current WMAP measurement of the age of the universe (13.7 Gyr; Spergel 
\etal\ 2003), it would be unrealistic for these galaxies to have extremely 
old central regions.  Dust likely contributes to the red colors, but they 
could also result from erroneous seeing measurements.  In the BdJ00 sample 
3 very late-type (Sdm/Irr) galaxies (see left panel of 
Fig.~\ref{fig:gradsK_Vrot}) and the extended tails of another 3 galaxies lie 
blue-ward of the model grids in their $R-K$ colors.  These low surface 
brightness galaxies may suffer from calibration andsky uncertainties couple
with the weakness of the model 
predictions at such low metallicities.  Finally, the BdJ00 sample has 3 
early-type galaxies whose entire profiles 
lie red-ward of the grids in their $B-R$ colors. Again, extremely old ages are
improbable, but significant dust absorption could contribute to the red colors 
of these three galaxies.  Thus, unlike some of the studies cited in 
\se{dust}, these observations remind us of our fragile understanding of dust
effects and the challenge we face in trying to separate them from stellar 
population effects.

\section{Results}\label{sec:results}

\subsection{Local and Global Trends in Age and Metallicity}\label{sec:global}

Figure~\ref{fig:AZ_mu} shows the local average age (left) and metallicity 
(right) as a function of local surface brightness.
There is a clear trend in the local age and metallicity as a function of local 
SB in the sense that regions of higher SB are older and more metal rich, but 
with large scatter, as also found in BdJ00.  This correlation suggests that 
the local potential plays an important role in the SFH.  

Figure~\ref{fig:Aeff} shows an effective average age, 
$\langle A \rangle_{\mathrm{eff}}$, for a galaxy, taken as the measured mean 
age in the 0.5--1.5$h$ averaged radial bin, as a function of central surface 
brightness (left plot) and total galaxy magnitude (right plot).  There
is a clear trend in $\langle A \rangle_{\mathrm{eff}}$ as a function of both
$\mu_{0,H_{\mathrm{m}},K}$ and $M_{H_{\mathrm{m}},K}$ with more luminous and 
higher CSB galaxies having older $\langle A \rangle_{\mathrm{eff}}$ 
(as expected from Fig.~\ref{fig:AZ_mu}).  However, in both cases, the roughly 
linear increasing trend at $\mu_{0,H_{\mathrm{m}},K} \ga 18.5$ 
mag arcsec$^{-2}$ and $M_{H_{\mathrm{m}},K} \ga -22.5$  mag seem to flatten for
brighter values.  For  $M_{H_{\mathrm{m}},K} \la -22.5$ there is large scatter 
and a much weaker (if any) trend of increasing age with total magnitude (but 
still containing the oldest galaxies).  
In Figure~\ref{fig:Aeff_hVrot} we plot $\langle A \rangle_{\mathrm{eff}}$ 
as a function of $h_{H,K}$ (left) and $V_{\mathrm{rot}}$ (right).
As expected from the correlation of $M_{H_{\mathrm{m}},K}$ with 
$V_{\mathrm{rot}}$ (Fig.~\ref{fig:galpars}), a trend of increasing 
$\langle A \rangle_{\mathrm{eff}}$ with $V_{\mathrm{rot}}$  is detected 
with a similar 
turnover in the slope of the correlation at higher rotational velocities 
($V_{\mathrm{rot}} \ga 120$ \kms, the location of the vertical dotted
line).  The left panel of Figure~\ref{fig:Aeff_hVrot} may show a 
weak trend of increasing age with $h$, however, the scatter is very large.
Note also that the spread in age is larger for smaller $h$ galaxies, and
reduces with increasing $h$.

Figures~\ref{fig:Zeff} \& \ref{fig:Zeff_hVrot} similar trends for the 
effective metallicity, $\log(Z_{\mathrm{eff}}/Z_{\odot})$ versus
$\mu_{0,H_{\mathrm{m}},K}$, $M_{H_{\mathrm{m}},K}$, and 
$V_{\mathrm{rot}}$, but no trends are seen with $h$.  
Effective 
metallicity increases with total magnitude up to 
$M_{H_{\mathrm{m}},K} \sim -22$, but likely saturates
for brighter galaxies at roughly solar metallicity (with large scatter).  
Similarly, in the right panel of Figure~\ref{fig:Zeff_hVrot}, 
$\log(Z_{\mathrm{eff}}/Z_{\odot})$ increases with $V_{\mathrm{rot}}$ up to 
$\sim 120$ \kms\ and then levels off with smaller scatter.  There is
no dependence of $\log(Z_{\mathrm{eff}}/Z_{\odot})$ with $h$.

\subsection{Global Trends in Age and Metallicity Gradients}
\label{sec:globgrads}

One of the conclusions from BdJ00 is that the amplitude of the age gradients 
increases from HSB to LSB galaxies.  However, this is likely an artifact of 
their linear gradient fitting technique out to different number of radial bins 
(scaled by the galaxy disk scale length), anywhere from 2--4 bins (\ie\ out 
to 1.5--3.5 scale lengths).  Assuming that the LSB galaxy photometry generally
does not extend to as many scale lengths as the HSB photometry, non-linear 
gradients in the galaxies (\eg\ steeper in the inner regions, flattening at 
large radii), could mimic a trend with SB.  To demonstrate this, in 
Figures~\ref{fig:Age_r} \& \ref{fig:Z_r}, we plot average age, 
$\langle A \rangle$, and metallicity,  $\log(Z/Z_{\odot})$, respectively, as a 
function of radius.  In each case, the left panels give radius in terms of 
disk scale lengths, and the right panels have the physical radius in kpc.  
Modulo fairly large uncertainties, we see that gradients are not linear over 
the extent of the galaxy.  The inner and outer slopes are often quite 
different, sometimes even changing sign.  Also, the galaxies are not all 
measured out to the same
number of scale lengths.  When plotted against number of scale lengths, most 
galaxies have steeper (negative) inner gradients that flatten off, or
even turn around, in the outer parts.  With the exception of extremely
early or late-types, the outer slopes are quite similar for most galaxies.  
However, when plotted as a function of physical scale (kpc), the slopes can be 
quite different, with a rough trend for the earliest types being steepest, and 
leveling off for later-types. 

Figures~\ref{fig:Age_r} \& \ref{fig:Z_r} demonstrate that fitting gradients 
out to a different number of scale lengths yields misleading results.  
In general, the larger the fit baseline, the shallower the measured 
gradient.  We verify this assertion in Figure~\ref{fig:dA_mu0} where we plot 
age gradients fit out to different numbers of radial bins for all the 
galaxies as a function of CSB (0--1.5$h_{H,K}$ (top left), 
0--2.5$h_{H,K}$ (top right),
0--3.5$h_{H,K}$ (bottom left), and 0--4.5$h_{H,K}$ (bottom right)).
Indeed, fitting fewer bins results in 
systematically larger (usually negative, but occasionally positive) age 
gradients. The more extended the baseline, the flatter the gradient.
LSB galaxy profiles do not extend as far as HSBs (in scale 
lengths, as indicated by the lower CSB galaxies dropping out of the longer 
baseline fits) and thus appear to have larger gradients.

Given the dangers of fitting gradients out to different radial extents, in 
the remainder of this analysis we consider ``inner'' and ``outer'' gradients
fitted over the 1$^{st}$--2$^{nd}$ radial bins and 
2$^{nd}$--4$^{th}$ bins, respectively\footnote{Note that these two 
definitions cannot be interpreted as bulge and 
disk gradients as the inner fit goes out to 1.5 disk scale lengths, and bulge 
scale lengths are typically $\sim$0.13 $h_d$ (Paper I).}.
This restriction greatly reduces the available galaxy parameter range, 
especially for the LSB galaxies.  Also note that inner gradients will be much 
more affected by dust (if present) than outer gradients.

\subsection{Age gradients}\label{sec:age}

In Figure~\ref{fig:dA_bd} we plot the inner (left panels) and outer (right 
panels) gradients measured in disk scale lengths (upper panels) and kpc (lower 
panels) as a function of CSB (left plot) and total magnitude (right plot).
Unlike BdJ00, we do not see a trend of d$\langle A \rangle/$dr (Gyr/$h$)
(inner or outer) with CSB.  Note that the low CSB galaxies are missing
in the outer gradient plots so one cannot assess any trend for the outer 
gradients with great confidence.  The current data suggest that outer 
gradients are generally smaller than those within.  When age gradients are 
plotted as a function of kpc, the inner gradients are slightly
steeper for higher CSB galaxies ($\mu_{0,H_{\mathrm{m}},K} \la 17.5$).  
The 3--4 outliers at high positive gradients are the S0 galaxies whose color 
gradients are inverted (bluer inward).  Such anomalous gradients have been 
seen before in S0 galaxies (\eg\ Emsellem \etal\ 2002) and are often 
interpreted as the result of a recent gaseous accretion followed by (central)
star 
formation.  These are also the few galaxies that deviate to much younger ages 
from the general trend of decreasing effective age with galaxy type (see left 
plot of Fig.~\ref{fig:AZeff_morph}) which would agree with the above 
interpretation of recent accretion and SF. 

There is a slight trend (with large scatter) for more luminous galaxies to 
have steeper d$\langle A \rangle/$dr (Gyr/$h$) negative gradients 
(inner and outer), and the fainter galaxies extending to positive 
gradients.  However, this trend disappears when the gradients are 
plotted as a function of kpc.  In this scale the outer gradients are greatest
(negative or positive) for the fainter galaxies, and become negligible for
the brighter ones.  A similar, somewhat stronger trend of 
steepening gradient (in scale lengths) with increasing disk scale length 
is seen in Figure~\ref{fig:dA_hVrot} (left).  The trend in 
d$\langle A \rangle/$dr (Gyr/kpc) is reversed (ignoring the 3--4 steep positive
gradient points) such that bigger galaxies have smaller gradients.  
This may reflect the larger galaxies as being less likely to have had any 
recent activity/interactions, as
the small gradients combined with the older ages for the bigger galaxies 
suggest an overall old age for the largest galaxies.  Similar trends
are seen with $V_{\mathrm{rot}}$ (as expected from the correlations between
$V_{\mathrm{rot}}$ with $h$ and $M$ in Fig.~\ref{fig:galpars}), but they are 
weaker (largely because of the smaller sample).

\subsection{Metallicity gradients}\label{sec:metal}

Figures~\ref{fig:dlogZ_bd} \& \ref{fig:dlogZ_hVrot} recast the information 
presented in Figures~\ref{fig:dA_bd} \& \ref{fig:dA_hVrot} discussed above, 
but now in terms of metallicity gradients.  

Metallicity gradients are less well measured and are more sensitive to dust
effects than age gradients (according to the triplex dust models discussed in
\se{dust}).  Nevertheless, the trends in the metallicity 
gradients with galaxy parameters are very similar to those with age gradients 
(see \S~\ref{sec:age}).  The trends in the magnitudes of the gradients as a 
function of galaxy parameters are comparable except the correlation changes 
sign.  This suggests a direct anti-correlation between age and metallicity 
gradients within a galaxy which is seen in Figure~\ref{fig:dlogZ_dA} (although 
with some scatter), such that stronger age gradients are associated
with weaker metallicity gradients (and vice versa).

\section{Discussion}\label{sec:discussion}

Our main findings can be summarized as follows:
\begin{description}
\item[(i)  ]  Our relative age determinations are robust under the assumption 
      that the underlying SFR of all disk galaxies is similar, and that no 
      major star bursts have occurred in the past 1--2 Gyr (which is justified
      on the basis of Kauffmann \etal\ (2003)).
\item[(ii) ] Dust cannot be ruled out as a contributor to color gradients in 
      spiral galaxies, although it is unlikely that the gradients are 
      largely due to dust extinction.
\item[(iii)] Age and metallicity correlate strongly with local surface 
      brightness: higher SB regions tend to be older and more metal-rich 
      (see Fig.~\ref{fig:AZ_mu}).  This indicates that the local 
      potential plays a significant role in the SFH of spiral galaxies.
\item[(iv) ] Age and metallicity, measured at an effective radius of 1$h$, 
      increase 
      with earlier Hubble type, $M_{H_{\mathrm{m}},K}$, $V_{\mathrm{rot}}$, 
      and $\mu_{0,H_{\mathrm{m}},K}$ but the trends flatten for $T \la 4$, 
      $M_{H_{\mathrm{m}},K} \la -22.5$ mag,  $V_{\mathrm{rot}} \ga 120$ 
      \kms, and $\mu_{0,H_{\mathrm{m}},K} \la 18.5$ mag arcsec$^{-2}$ (see 
      Figs.~\ref{fig:Aeff}--\ref{fig:Zeff_hVrot} \& \ref{fig:AZeff_morph}).  
      Age also correlates
      weakly with scale length (with scatter decreasing
      with galaxy size, see right plot of Fig.~\ref{fig:Aeff_hVrot}).
\item[(v)  ] Age and, to a lesser extent, metallicity gradients show radial 
      structure, with generally steeper gradients in the inner parts of the 
      galaxy. Care must thus be taken when defining the gradient fit region
      (see Figs.~\ref{fig:Age_r}--\ref{fig:dA_mu0}).
\item[(vi) ] Age gradients, as measured as a function of scale length, show
      correlations with luminosity, size, and rotational velocity 
      (Fig.~\ref{fig:dA_bd}), especially in the inner disk.  Trends in
      metallicity gradients with galaxy parameters are weaker 
      (Figs.~\ref{fig:dlogZ_bd} \& \ref{fig:dlogZ_hVrot}).
\item[(vii)] Age gradients do not correlate with CSB, contrary to the findings 
      of BdJ00, which we attribute to inconsistencies in their fit radii
      (see left plot of Fig.~\ref{fig:dA_bd}).  This is not at odds with 
      statement (iii) since $\mu_0$ and $h$ are not strongly correlated.
\end{description}
These observations are consistent with a picture where (i) higher surface 
brightness regions of galaxies formed their stars earlier 
than lower surface brightness regions, or at a similar
epoch but on shorter timescales,  and (ii) the SFHs at a given SB level,
which lead to age gradients, are modulated by the overall potential of the 
galaxy such that brighter/higher rotational velocity galaxies formed earlier.  
These trends reach saturation for the brightest and highest CSB galaxies.  

An earlier formation time for more massive galaxies is clearly in conflict 
with hierarchical galaxy formation which predicts that more massive galaxies
form late.  This discrepancy could be remedied if there is a mechanism at
work that prevents the gas in lower mass galaxies from cooling and 
forming stars at early times.  Feedback processes are often invoked as a 
possible solution, but no prescription has been found thus far that
agrees with all observational constraints (\eg\ Kauffmann \etal\ 2003;
Bell \etal\ 2003).

In a related manner, the semi-analytic models of hierarchical galaxy 
formation of Kauffman (1996) predict a correlation between bulge-to-disk 
ratio and luminosity-weighted age, such that the 
bulges of late-type spirals should be older (up to $\sim 4$ Gyr) than those of 
early-type spirals, although this correlation could be erased by any
significant inflow from the disk after the last major merger.  
Figure~\ref{fig:AZeff_morph} (left) shows a strong correlation of effective
age with galaxy type, but in the opposite sense of Kauffmann's prediction;
later-type galaxies have systematically younger effective ages. 
Clearly, 
mixing by a bar could  erase the predicted age trend.  If bars trigger 
radial flows the 
gradients of strongly barred galaxies would be flat (the flattening 
would occur on short enough timescales).  To look for such mixing effects, 
we plot age and metallicity gradients as a function of barredness in 
Figure~\ref{fig:dAZ_bar}.  
No trends are seen when the gradients are plotted 
in scale lengths, but against kpc, the strongly barred galaxies may have
smaller age gradients (inner and outer), though this observation is based
on small statistics.  This agrees with Martin \& Roy (1994) who found similar 
flattening of O/H metallicity gradients with bar strength (their trend
also disappears when the gradients are plotted in scale lengths).
Thus, under all assumptions, our observations cannot be reproduced by the 
Kauffmann (1996) models.  The same model deficiency responsible for the 
backwards age-size correlation, as compared against observations, could also
be the cause of the discrepancy in bulge morphology versus age seen in 
the Kauffmann (1996) models.

Prantzos \& Boissier (2000; hereafter PB00) present chemo-spectrophotometric 
models of spiral galaxy evolution.  They adopt a Schmidt-type law for the SFR, 
which is proportional to the gas surface density and varies with galactocentric
radius (dynamical time), 
assume an inflow rate (of unenriched gas) that decreases exponentially with 
time and increases with surface density and galaxy mass, and assume that the 
gas settles into an exponential disk (bulges are not modeled), but do not 
consider radial 
flows (\eg\ by viscosity which could create gradients, or due to a bar which 
could flatten radial gradients on small timescales).  The star formation 
efficiency and infall rate are free parameters tuned to match 
the Milky Way observational constraints.  The models are then extended to 
other disk galaxies by adopting the Cold Dark Matter(CDM)-based scaling 
relations of Mo, Mao, \& White (1998).  As such, their galaxy disk radial 
profiles are fully described by just two parameters: rotational velocity, 
$V_{\mathrm{rot}}$ (assuming a constant disk-to-halo mass ratio), and the halo 
spin  parameter, $\lambda$.  A third parameter describing the formation 
redshift would be required for a description fully consistent with the
CDM hierarchical models of galaxy formation but, as of yet, there is no clear
definition for the time of formation for an individual galaxy.  Hence, PB00 
assume that all galaxies started forming at the same epoch (13.5 Gyr ago) 
but evolve at different rates.  

The PB00 models give predictions for O/H abundances and gradients that 
as could be measured in bright \ion{H}{2} regions in nearby galaxies.  Note 
that O/H determinations in \ion{H}{2} regions probe the 
present-day ISM abundances and are insensitive to abundance evolution with 
time.  Still, their model predictions and observations of O/H gradients 
can be compared, at least indirectly, to our stellar luminosity weighted 
gradients.  PB00 find that the absolute central abundance for a given 
$\lambda \ge 0.03$
correlates with $V_{\mathrm{rot}}$ (and total magnitude) 
such that faster rotators have larger central abundances, but the trend 
saturates for $V_{\mathrm{rot}} \ga 220$ \kms.  Furthermore, at a 
given $V_{\mathrm{rot}}$, the central abundance decreases with increasing 
$\lambda$ but again saturates for galaxies with $V_{\mathrm{rot}} \ga 220$ 
\kms\ above which central abundances are high, regardless of 
$\lambda$.  By extension, we should see a trend of increasing central
abundance with $V_{\mathrm{rot}}$, with significant spread (larger at lower
$V_{\mathrm{rot}}$) due to different
values of  $\lambda$, but this trend would flatten and show less dispersion
above $V_{\mathrm{rot}} \ga 220$ \kms.  
This is roughly what Figures~\ref{fig:Zeff} \&~\ref{fig:Zeff_hVrot} show
for the effective metallicity at the half-light radius versus 
M$_{H_{\mathrm{m}},K}$ and $V_{\mathrm{rot}}$ (central values are too 
uncertain due to likely higher concentrations of dust and seeing mismatches, 
and PB00's models do not consider bulges).  Unfortunately, our galaxies do not
exceed $V_{\mathrm{rot}} \sim 250$ \kms, but the general trend above,
including the decrease in scatter with $V_{\mathrm{rot}}$, is confirmed.

The right side of Figure~\ref{fig:dlogZ_bd} shows our measured metallicity 
gradients as a function of total galaxy magnitude; gradients in disk scale 
lengths are shown in the upper panels while gradients in kpc are shown 
underneath.  The left and right panels show the ``inner'' and ``outer''
gradients respectively.  When plotted against scale length, the inner 
or outer gradients show no clear trends.  However, when
plotted in kpc, a trend emerges such that brighter galaxies have smaller
gradients.
Since scale length increases with luminosity (see Fig.~\ref{fig:galpars}), 
the correlation goes away when the gradients are measured per disk scale 
length.  If this effect is real, the disappearance of a trend of metallicity 
gradients in units of scale length may suggest a self-similar pattern in disk 
galaxies.  As suggested by 
Combes (1998), a ``universal'' slope per disk scale length might be 
explained by the viscous disk models of Lin \& Pringle (1987), although 
model predictions are not conclusive as of yet.  
The Garnett \etal\ (1997) compilation of O/H gradients shows
a similar signature, where abundance gradients in dex/kpc are steeper and
exhibit greater scatter for lower luminosity disks, but this trend goes away 
when gradients are expressed in dex/$h$.  Note, however, that the abundance 
gradients compiled in van Zee \etal\ (1998) and reported in PB00 
do show a clear increase with magnitude when plotted as dex/$h$, thus 
thwarting the interpretation of a universal abundance gradient per scale 
length.

We have measured compelling trends for the effective age and 
metallicity in spiral galaxies and their inner and outer disk
gradients as a function of surface brightness, luminosity, 
rotational velocity, and size.  These trends are not well reproduced 
by current semi-analytical models of galaxy evolution but will undoubtedly 
serve as effective constraints for future models.
An important limitation inherent in our color-based analysis is its
sensitivity to dust extinction which could mimic gradients in age and,
particularly, metallicity.  A spectroscopic analysis of radially resolved
line indices of spiral galaxies is underway in order to alleviate this
problem.

\appendix

\section{\ion{H}{1} Line widths} \label{HIwidths}

\ion{H}{1} line widths were collected from various sources in the literature; 
Courteau (1997) [7gals]; Bottinelli \etal\ (1990) [74 gals]; 
Theureau \etal\ (1998) [11 gals]; Haynes \etal\ (1999) [1 gal]; 
de~Blok \& Bosma (2002) [1 gal]; de~Blok, McGaugh, \& Rubin (2001) [4 gals].  
For the Bottinelli \etal\ (1990) and  Theureau 
\etal\ (1998) samples, the line width at 20\% and 50\% of the peak flux, $W50$ 
and $W20$, were provided, though not for all galaxies.  In cases where only
$W20$ was available, a conversion to $W50$ was made from a linear 
least-squares fit between these two quantities derived from those for galaxies 
for which both measurements were available.  
The conversions between the fully corrected line widths are as follows:

For the Bottinelli \etal\ (1990) sample: $W50 = 0.98*W20 - 19.21 \,\,[N = 3325]$

For the Theureau \etal\ (1998) sample: $W50 = 1.00*W20 - 17.01 \,\,[N = 2055]$

Haynes \etal\ (1999) give a different measure of the line width
denoted $W21$ and defined as the full width between the velocity channels
at the 50\% level of each horn.
We can convert the Haynes \etal\ (1999)'s $W21$ to Bottinelli \etal\ 
(1990)'s $W50$ as:
$W50(Bot) = 1.02*W21(Haynes) + 1.39 \,\,[N = 275]$,

and Courteau (1997)'s $V_{\mathrm{max}}$ and Botinelli \etal\ 
(1990)'s $W50$ are matched with:

$W50(Bot) = 0.99*V_{\mathrm{max}} - 9.31 \,\,[N=122]$

For 4 of the LSBs in the BdJ00 sample we obtained $V_{\mathrm{max}}$ values 
from de~Blok, McGaugh, \& Rubin (2001) and de~Blok \& Bosma (2002).
A comparison between the Courteau (1997) $V_{\mathrm{max}}$ values and the 
Bottinelli \etal\ (1990) $W50$s for galaxies in common reveals little 
difference, and we assume here that $V_{\mathrm{max}}$ values for the LSBs 
also map directly into $W50$.  

Multiple measurements of a galaxy from differnet samples vary
typically by less than 10\%, more than accurate enough for our
purposes.  In order to ensure accurate V$_{\mathrm{rot}}$ measurements, we 
restricted ourselves to galaxies with inclinations greater
than 35$\degr$.  Our final sample has 98 galaxies with reliable 
rotational velocity measurements.

\clearpage
% Fig. 1
\begin{figure}
\vspace{2cm}
%\plotone{f1.eps}
%\plotone{figures/bw_galpars.eps}
\vspace{0.5cm} \caption{Correlation between galaxy parameters: rotational velocity, 
            $V_{rot}$ (kpc), central surface brightness 
            $\mu_{0,H_{\mathrm{m}},K}$, 
            total magnitude, $M_{H_{\mathrm{m}},K}$, and disk scale length 
            $h_{H,K}$.  The dotted line is at 
            log$_{10}$($V_{\mathrm{rot}}$ = 120 \kms).  
            Points with dotted 
            error bars represent the galaxies from our sample.  Points with
            solid error bars are the BdJ00 sample.
    \label{fig:galpars}}
\end{figure}
% Fig. 2
\begin{figure}
\vspace{2cm}
%\plotone{f2.eps}
%\plotone{figures/bw_pars_morph.eps}
\vspace{0.5cm} \caption{Correlation between galaxy parameters and morphological type. 
             The dotted line is at log$_{10}$($V_{\mathrm{rot}}$ = 120 \kms).
    \label{fig:pars_morph}}
\end{figure}
% Fig. 3
\begin{figure}
\vspace{2cm}
%\plottwo{f3a.eps}{f3b.eps}
%\plottwo{figures/gx_tracks_salp_chab.eps}{figures/gx_peg_tracks.eps}
\vspace{0.5cm} \caption{Comparison of the stellar population model tracks for different
             initial mass functions; the Salpeter (1955) IMF 
             (eq.~\ref{eq:salp}), and the Chabrier (2003) IMF 
             (eq.~\ref{eq:chab}) for the GALAXEV models (left
             panel) and of the Salpeter versus the Kroupa (2001) 
             (eq.~\ref{eq:kroupa01}) IMFs for the
             PEGASE models (right panel).  The GALAXEV models are also 
             plotted on the right panel for comparison between two
             different SSP models.
\label{fig:tracks_imf}}
\end{figure}
% Fig. 4
\begin{figure}
\vspace{2cm}
%\plotone{f4.eps}
%\plotone{figures/sfr_time.eps}
\vspace{0.5cm} \caption{Time evolution for the exponential (eq.~\ref{eq:exp}), upper 
             panel, and Sandage (eq.~\ref{eq:sandage}), lower panel, star 
             formation histories (solid curves).  The dotted curve is 
             a Sandage-style burst of star formation in which 10\% of the
             total mass of stars are formed.  
             See Figure~\ref{fig:tracks_burst2} 
             for the effect of such a burst on the population model grids.
\label{fig:sfr_time}}
\end{figure}
\clearpage
% Fig. 5
\begin{figure}
\vspace{2cm}
%\plotone{f5.eps}
%\plotone{figures/sand_exp.eps}
\vspace{0.5cm} \caption{Comparison of the stellar population model tracks for exponential
             (eq.~\ref{eq:exp}) and Sandage (eq.~\ref{eq:sandage}) star 
             formation histories. 
\label{fig:tracks_sfh}}
\end{figure}
\clearpage
% Fig. 6
\begin{figure}
\vspace{2cm}
%\plottwo{f6a.eps}{f6b.eps}
%\plottwo{figures/gx_burst_1Gyr_0.1.eps}{figures/gx_burst_1Gyr_0.5.eps}
\vspace{0.5cm} \caption{Comparison of the stellar population model tracks for a simple 
             exponential SFH (eq.~\ref{eq:exp}) and one with a 1 Gyr old 
             Sandage (eq.~\ref{eq:sandage}) burst contributing 10\% of the 
             total mass (left plot) and 50\% of the mass (right plot). 
\label{fig:tracks_burst1}}
\end{figure}
% Fig. 7
\vspace{2cm}
\begin{figure}
%\plottwo{f7a.eps}{f7b.eps}
%\plottwo{figures/gx_burst_4Gyr_0.1.eps}{figures/gx_burst_4Gyr_0.5.eps}
\vspace{0.5cm} \caption{Comparison of the stellar population model tracks for a simple
             exponential SFH (eq.~\ref{eq:exp}) and one with a 4 Gyr old 
             Sandage (eq.~\ref{eq:sandage}) burst contributing 10\% of the 
             total mass (left plot, and see Figure~\ref{fig:sfr_time}) 
             and 50\% of the mass (right plot). 
\label{fig:tracks_burst2}}
\end{figure}
%\clearpage
% Fig. 8
\begin{figure}
%\plotone{f8.eps}
\vspace{-5cm}
%\plotone{figures/bulge_disk.eps}
     \caption{Disk colors (average of 1.5--2.5 disk scale length radial bin)
             as a function of bulge colors (average of 0.0--0.5 disk scale 
             length radial bin).  The solid horizontal lines represent a 
             one-to-one mapping (for reference only).  $H_{\mathrm{m}}$ is the 
             ``modified'' $H$-band magnitude for the Courteau \etal\ sample, 
             converted to $K$-band with 2MASS $H-K$ colors (see text for 
             details) for direct comparison
             with the $K$-band data of the BdJ00 sample.
    \label{fig:bulge_disk}}
\end{figure}
% Fig. 9
\begin{figure}
%\vspace{2cm}
%\plottwo{f9a.eps}{f9b.eps}
%\plottwo{figures/grads_morph1.eps0}{figures/grads_morph2.eps0}
\vspace{0.5cm} \caption{Near-IR--optical color--color plots for the Courteau \etal\
             sample separated by morphological
             type.  The galaxy centers are indicated by solid symbols: 
             circles for Type-I galaxies, triangles for Type-II, and asterisks
             for Transition galaxies.  Open circles denote the half-light 
             radius of the disk.  The line types for the galaxy color profiles
             represent bar strength: solid for bar-less (A), dashed for mild 
             bars (AB), and dot-dashed for strong bars (B).  Average errors
             due to the global calibration, as well as sky uncertainties for 
             the innermost and outermost points are shown as crosses. 
             Over-plotted are the Bruzual \& Charlot (2003; GALAXEV) stellar 
             population models for an exponential SFH at different
             metallicities.  Lines of constant average age
             $\langle A \rangle$ (see eq.\ref{eq:avgage}) are the dashed, 
             roughly vertical lines, while lines of constant metallicity 
             are solid and nearly horizontal.  In the upper left corner of
             each panel, triplex and foreground screen dust models are
             plotted.  For the triplex model, the Milky Way dust extinction
             curve and albedo from Gordon, Calzetti, \& Witt (1997) was
             adopted.  The galaxy center and disk half-light radius are 
             denoted by solid and open circles respectively.
     \label{fig:grads_morph}}
\end{figure}
\clearpage
% Fig. 10
\begin{figure}
\vspace{2cm}
%\plottwo{f10a.eps}{f10b.eps}
%\plottwo{figures/grads_HI1.eps0}{figures/grads_HI2.eps0}
\vspace{0.5cm} \caption{Near-IR--optical color--color plots for the Courteau \etal\
             sample separated by rotational velocity, 
             $V_{\mathrm{rot}}$ (\kms).  Symbols and line-types are as 
             in Figure~\ref{fig:grads_morph}.
    \label{fig:grads_Vrot}}
\end{figure}
% Fig. 11
\begin{figure}
\vspace{2cm}
%\plottwo{f11a.eps}{f11b.eps}
%\plottwo{figures/gradsK_HI1.eps}{figures/gradsK_HI2.eps}
\vspace{0.5cm} \caption{Near-IR--optical color--color plots separated by rotational
             velocity, $V_{\mathrm{rot}}$ (\kms), for the BdJ00 sample.  
    \label{fig:gradsK_Vrot}}
\end{figure}
\clearpage
% Fig. 12
\begin{figure}
\vspace{2cm}
%\plottwo{f12a.eps}{f12b.eps}
%\plottwo{figures/grads_mag2.eps0}{figures/grads_mag1.eps0}
\vspace{0.5cm} \caption{Near-IR--optical color--color plots for the Courteau \etal\
             sample separated by H-band magnitude,
             $M_{H}$.  Symbols and line-types are as 
             in Figure~\ref{fig:grads_morph}. 
    \label{fig:grads_Mh}}
\end{figure}
% Fig. 13
\begin{figure}
\vspace{2cm}
%\plottwo{f13a.eps}{f13b.eps}
%\plottwo{figures/grads_sb2.eps0}{figures/grads_sb1.eps0}
\vspace{0.5cm} \caption{Near-IR--optical color--color plots for the Courteau \etal\
             sample separated by H-band central
             surface brightness, $\mu_{0,H}$.  Symbols and line-types are as 
             in Figure~\ref{fig:grads_morph}. 
    \label{fig:grads_sb}}
\end{figure}
% Fig. 14
\begin{figure}
\vspace{5cm}
%\plottwo{f14a.eps}{f14b.eps}
%\plottwo{figures/A_muHK.eps}{figures/Z_muHK.eps}
\vspace{0.5cm} \caption{Average age, $\langle A \rangle$ (left), and metallicity, 
             log$_{10}(Z/Z_{\odot})$ (right) as a function of local
             $H_{\mathrm{m}}$ and $K$-band surface brightness, where 
             $H_{\mathrm{m}}$ is the ``modified'' $H$-band SB for the 
             Courteau \etal\ sample, converted to $K$-band 
             with 2MASS $H-K$ colors (see text for details) for direct 
             comparison with the $K$-band data of the BdJ00 sample.  The 
             $H_{\mathrm{m}}$ data is distinguished with dotted error bar 
             lines. The different symbols represent total 
             $H_{\mathrm{m}},K$-band magnitude bins for each galaxy.
             The dotted lines in the metallicity plot (right) denote the
             limits of the population model grids.  Points at or near these 
             limits for metallicity, and those at 12.9 Gyr for the average 
             age should be interpreted with caution.
    \label{fig:AZ_mu}}
\end{figure}
\clearpage
% Fig. 15
\begin{figure}
\vspace{2cm}
%\plottwo{f15a.eps}{f15b.eps}
%\plottwo{figures/bw_Aeff_mu0.eps}{figures/bw_Aeff_M.eps}
\vspace{0.5cm} \caption{``Effective'' average age (average of 0.5--1.5$h$ bin), 
             $\langle A \rangle_{\mathrm{eff}}$ 
             as a function of central surface brightness (left) and
             total $H_{m},K$-band galaxy magnitude (right) for all 158 
             galaxies. See caption of Figure~\ref{fig:AZ_mu} for the 
             definition of $H_{\mathrm{m}}$. 
    \label{fig:Aeff}}
\end{figure}
% Fig. 16
\begin{figure}
\vspace{2cm}
%\plottwo{f16a.eps}{f16b.eps}
%\plottwo{figures/bw_Aeff_h.eps}{figures/bw_Aeff_Vrot.eps}
\vspace{0.5cm} \caption{``Effective'' average age as a function of scale length
             (left) for all 158 galaxies and rotational velocity (right),
             $V_{\mathrm{rot}}$ for all 98 galaxies for which we have
             reliable $V_{\mathrm{rot}}$ measurements.
    \label{fig:Aeff_hVrot}}
\end{figure}
\clearpage
% Fig. 17
\begin{figure}
\vspace{2cm}
%\plottwo{f17a.eps}{f17b.eps}
%\plottwo{figures/bw_Zeff_mu0.eps}{figures/bw_Zeff_M.eps}
\vspace{0.5cm} \caption{``Effective'' metallicity 
             log$_{10}(Z_{\mathrm{eff}}/Z_{\odot})$  as a function of 
             central surface brightness (left) and total 
             $H_{\mathrm{m}},K$-band galaxy magnitude (right)
             for all 158
             galaxies.  See caption of Figure~\ref{fig:AZ_mu} for the 
             definition of $H_{\mathrm{m}}$.
    \label{fig:Zeff}}
\end{figure}
% Fig. 18
\begin{figure}
\vspace{2cm}
%\plottwo{f18a.eps}{f18b.eps}
%\plottwo{figures/bw_Zeff_h.eps}{figures/bw_Zeff_Vrot.eps}
\vspace{0.5cm} \caption{``Effective'' metallicity,
             log$_{10}(Z_{\mathrm{eff}}/Z_{\odot})$,
             as a function of scale 
             length (left) for all 158 fitted galaxies, and 
             rotational velocity (right),
             $V_{\mathrm{rot}}$ for all 98 galaxies for which we have
             reliable $V_{\mathrm{rot}}$ measurements. 
    \label{fig:Zeff_hVrot}}
\end{figure}
\clearpage
% Fig. 19
\begin{figure}
\vspace{2cm}
%\plottwo{f19a.eps}{f19b.eps}
%\plottwo{figures/Age_r.eps}{figures/Age_r_kpc.eps}
\vspace{0.5cm} \caption{Age as a function of radius for all galaxies.
             The left plot gives the radius in terms of the measure scale 
             length $h_{H,K}$ while the right plot gives the radius in
             physical units (kpc).  Dotted lines represent our sample whereas
             solid lines are the BdJ00 sample.
    \label{fig:Age_r}}
\end{figure}
% Fig. 20
\begin{figure}
\vspace{2cm}
%\plottwo{f20a.eps}{f20b.eps}
%\plottwo{figures/logZ_r.eps}{figures/logZ_r_kpc.eps}
\vspace{0.5cm} \caption{Same as Figure~\ref{fig:Age_r} but for metallicity.
    \label{fig:Z_r}}
\end{figure}
\clearpage
% Fig. 21
\begin{figure}
\vspace{2cm}
%\plotone{f21.eps}
%\plotone{figures/bw_dA_mu0HK.eps}
\vspace{0.5cm} \caption{Age gradient as a function of central surface brightness,
            $\mu_{0,H_{\mathrm{m}},K}$ for all galaxies.  The gradients 
            are fit out to a different number of radial bins: 
            0--1.5 $h_{H,K}$ (top left), 0--2.5 $h_{H,K}$ (top right),
            0--3.5 $h_{H,K}$ (bottom left), and 0--4.5 $h_{H,K}$ 
            (bottom right).
    \label{fig:dA_mu0}}
\end{figure}
\clearpage
% Fig. 22
\begin{figure}
\vspace{2cm}
%\plottwo{f22a.eps}{f22b.eps}
%\plottwo{figures/bw_dA_mu0HK_bd.eps}{figures/bw_dA_MHK_bd.eps}
\vspace{0.5cm} \caption{Age gradient as a function of $H_{\mathrm{m}}$,$K$-band central 
            surface brightness, $\mu_{0,H_{\mathrm{m}},K}$ (left) and total 
            magnitude, $M_{H_{\mathrm{m}},K}$ (right) for all galaxies.  The 
            fit ranges represent ``inner''gradients (left panels) fit out 
            to 1.5 $H$,$K$-band disk scale lengths, $h_{H,K}$, and 
            ``outer'' gradients (right panels) fit from 1.5 to 3.5 $h_{H,K}$.
            The upper and lower plots show gradients as a function of scale 
            length and kpc respectively.
    \label{fig:dA_bd}}
\end{figure}
% Fig. 23
\begin{figure}
\vspace{2cm}
%\plottwo{f23a.eps}{f23b.eps}
%\plottwo{figures/bw_dA_hHK_bd.eps}{figures/bw_dA_Vrot_bd.eps}
\vspace{0.5cm} \caption{Same as Figure~\ref{fig:dA_bd} except for disk scale lengths,  
             log$(h_{H,K})$ (left), and rotational velocity, 
             $V_{\mathrm{rot}}$ (right).
    \label{fig:dA_hVrot}}
\end{figure}
\clearpage
% Fig. 24
\begin{figure}
\vspace{2cm}
%\plottwo{f24a.eps}{f24b.eps}
%\plottwo{figures/bw_dlogZ_mu0HK_bd.eps}{figures/bw_dlogZ_MHK_bd.eps}
\vspace{0.5cm} \caption{Same as Figure~\ref{fig:dA_bd} except for metallicity gradients. 
    \label{fig:dlogZ_bd}}
\end{figure}
% Fig. 25
\begin{figure}
\vspace{2cm}
%\plottwo{f25a.eps}{f25b.eps}
%\plottwo{figures/bw_dlogZ_hHK_bd.eps}{figures/bw_dlogZ_Vrot_bd.eps}
\vspace{0.5cm} \caption{Same as Figure~\ref{fig:dA_hVrot} except for metallicity 
             gradients.
    \label{fig:dlogZ_hVrot}}
\end{figure}
% Fig. 26
\begin{figure}
%\plotone{f26.eps}
%\plotone{figures/bw_dlogZ_dA_bd.eps}
    \vspace{-6cm}
\vspace{0.5cm} \caption{Metallicity gradient as a function of age gradient for 
             all 158 galaxies.
    \label{fig:dlogZ_dA}}
\end{figure}
% Fig. 27
\begin{figure}
%\plottwo{f27a.eps}{f27b.eps}
%\plottwo{figures/bw_Aeff_morph.eps}{figures/bw_Zeff_morph.eps}
\vspace{0.5cm} \caption{Effective age (left) and metallicity (right) as a 
             function of morphological type for all 158 galaxies.
    \label{fig:AZeff_morph}}
\end{figure}
\clearpage
% Fig. 28
\begin{figure}
\vspace{2cm}
%\plottwo{f28a.eps}{f28b.eps}
%\plottwo{figures/bw_dA_morph_bd.eps}{figures/bw_dlogZ_morph_bd.eps}
\vspace{0.5cm} \caption{Age gradient (left) and metallicity gradient (right) as a 
             function of morphological type for all 158 galaxies.
             The upper panels plot radius in terms of the measured scale 
             length $h_{H,K}$ while the bottom panels plot 
             radius in physical units (kpc).
    \label{fig:dAZ_morph}}
\end{figure}
% Fig. 29
\begin{figure}
\vspace{2cm}
%\plottwo{f29a.eps}{f29b.eps}
%\plottwo{figures/Abar_hst.eps}{figures/Zbar_hst.eps}
\vspace{0.5cm} \caption{Age gradient (left) and metallicity gradient (right) 
             histograms as a function of barredness for all 158 galaxies.
             The upper panels plot the radius in terms of the measured scale 
             length $h_{H,K}$ while the bottom panels plot the 
             radius in physical units (kpc).  The fit ranges represent 
             ``inner''gradients (left panels) fit out to 1.5 $h_{H,K}$, and 
             ``outer'' gradients (right panels) fit from 1.5 to
             3.5 $h_{H,K}$.  The vertical dotted lines are located at
             zero gradient for reference.
    \label{fig:dAZ_bar}}
\end{figure}

\end{document}